\documentclass[twocolumn]{aastex62}
\pdfoutput=1 
\usepackage{amsmath,amstext}
\usepackage[T1]{fontenc}
\usepackage{apjfonts} 

\usepackage{graphicx}
\usepackage{graphics}
\usepackage{color}
\definecolor{dartmouthgreen}{rgb}{0.05, 0.5, 0.06}
\usepackage[caption=false]{subfig}
\usepackage{natbib}

\def\arcsec{\hbox{$^{\prime\prime}$}}
\def\deg{\hbox{$^\circ$}}
\def\hr{\textsuperscript{h}}
\def\min{\textsuperscript{m}}
\def\sec{\textsuperscript{s}\hspace{-0.7mm}}

\def\nh{$N_{\rm H}$}

\def\init{\hspace{0.75 mm}}

\def\nh{N_\mathrm{H}}

\def\chandra{\textit{Chandra}}

\def\WISE{\textit{WISE}}
\def\nustar{\textit{NuSTAR}}

\graphicspath{ {Images/} }

\shorttitle{Triple AGN in a Mid-infrared Selected Merger}
\shortauthors{R.\init W.\init Pfeifle et al.}

\begin{document}

\title{A Triple AGN in a Mid-Infrared Selected Late Stage Galaxy Merger}

\author{Ryan W. Pfeifle}
\affiliation{Department of Physics and Astronomy, George Mason University, MS3F3, 4400 University Drive, Fairfax, VA 22030, USA}

\author{Shobita Satyapal}
\affiliation{Department of Physics and Astronomy, George Mason University,  MS3F3, 4400 University Drive, Fairfax, VA 22030, USA}

\author{Christina Manzano-King}
\affiliation{Department of Physics and Astronomy, University of California, Riverside, 900 University Avenue, Riverside, CA, 92521, USA}

\author{Jenna Cann}
\affiliation{Department of Physics and Astronomy, George Mason University,  MS3F3, 4400 University Drive, Fairfax, VA 22030, USA}

\author{Remington O. Sexton}
\affiliation{Department of Physics and Astronomy, University of California, Riverside, 900 University Avenue, Riverside, CA, 92521, USA}

\author{Barry Rothberg}
\affiliation{LBT Observatory, University of Arizona, 933 N. Cherry Ave., Tucson, AZ 85721, USA}
\affiliation{Department of Physics and Astronomy, George Mason University, MS3F3, 4400 University Drive, Fairfax, VA 22030, USA}

\author{Gabriela Canalizo}
\affiliation{Department of Physics and Astronomy, University of California, Riverside, 900 University Avenue, Riverside, CA, 92521, USA}

\author{Claudio Ricci}
\affiliation{N\'ucleo de Astronom\'ia de la Facultad de Ingenier\'ia, Universidad Diego Portales, Av. Ej\'ercito Libertador 441, Santiago, Chile}
\affiliation{Kavli Institute for Astronomy and Astrophysics, Peking University, Beijing 100871, China}

\author{Laura Blecha}
\affiliation{Department of Physics, University of Florida, P.O. Box 118440, Gainesville, FL 32611-8440, USA}

\author{Sara L. Ellison}
\affiliation{Department of Physics and Astronomy, University of Victoria, Victoria, BC V8P 1A1, Canada}

\author{Mario Gliozzi}
\affiliation{Department of Physics and Astronomy, George Mason University,  MS3F3, 4400 University Drive, Fairfax, VA 22030, USA}

\author{Nathan J. Secrest}
\affiliation{U.S. Naval Observatory, 3450 Massachusetts Avenue NW, Washington, DC 20392, USA}

\author{Anca Constantin}
\affiliation{Department of Physics and Astronomy, James Madison University, Harrisonburg, VA 22807, USA}

\author{Jenna B. Harvey}
\affiliation{Department of Physics and Astronomy, James Madison University, Harrisonburg, VA 22807, USA}

\correspondingauthor{Ryan W.\init Pfeifle}
\email{rpfeifle@masonlive.gmu.edu}

\begin{abstract}
The co-evolution of galaxies and the supermassive black holes (SMBHs) at their centers via hierarchical galaxy mergers is a key prediction of  $\Lambda$CDM cosmology. As gas and dust are funneled to the SMBHs during the merger, the SMBHs light up as active galactic nuclei (AGNs). In some cases, a merger of two galaxies can encounter a third galaxy, leading to a triple merger, which would manifest as a triple AGN if all three SMBHs are simultaneously accreting. Using high-spatial resolution X-ray, near-IR, and optical spectroscopic diagnostics, we report here a compelling case of an AGN triplet with mutual separations $<$10 kpc in the advanced merger SDSS J084905.51+111447.2 at $\rm{z}=0.077$. The system exhibits three nuclear X-ray sources, optical spectroscopic line ratios consistent with AGN in each nucleus, a high excitation near-IR coronal line in one nucleus\textcolor{black}{, and broad Pa$\alpha$ detections in two nuclei}. Hard X-ray spectral fitting reveals a high column density along the line of sight, consistent with the picture of late-stage mergers hosting heavily absorbed AGNs. Our multiwavelength diagnostics support a triple AGN scenario, and we rule out alternative explanations such as star formation activity, shock-driven emission, and emission from fewer than three AGN. The dynamics of gravitationally bound triple SMBH systems can dramatically reduce binary SMBH inspiral \textcolor{black}{timescales}, providing a possible means to surmount the ``Final Parsec Problem.'' AGN triplets in advanced mergers are the only observational forerunner to bound triple SMBH systems and thus offer a glimpse of the accretion activity and environments of the AGNs prior to the gravitationally-bound triple phase.

\end{abstract}


\section{Introduction}

Observational campaigns and theoretical studies have shown both that supermassive black holes (SMBHs) reside at the centers of most galaxies and that galaxy interactions are ubiquitous in the Universe. As a result, galaxies grow and evolve hierarchically through collisions (e.g. \citealp{toomre1972,schweizer1982, barnes1992,schweizer1996,hibbard1996,rothberg2004}), during which gravitational torques drive gas reservoirs toward the centers of each galaxy, potentially fueling the SMBHs at their centers \citep{barnes1992,mihos1996,hopkins2008a,blecha2018}. Recent simulations predict that late-stage galaxy mergers --- with nuclear pair separations $\lesssim$ 10 kpc --- facilitate the most rapid black hole growth and represent the merger stage during which both SMBHs are expected to begin accreting as active galactic nuclei (AGNs) \citep{hopkins2008a,blecha2018}. Such dual AGNs are predicted to be highly obscured by gas and dust, consequently exhibiting red mid-infrared colors \citep{blecha2018}. \textcolor{black}{Indeed, observational studies have demonstrated that late stage galaxy mergers (pair separations $<$ 10 kpc) and post-mergers host higher fractions of dust obscured AGNs than do isolated galaxies in rigorously matched control samples, and in fact mid-infrared selection identifies a larger quantity of obscured AGNs in mergers than traditional optical selection techniques \citep{satyapal2014,weston2017,ellison2019}. Despite their important role in black hole and galaxy evolution, however, dual AGNs are rare systems, and less than 30 have been robustly verified in the literature (see Table 8 in \citealp{satyapal2017} and references therein). The rarity of dual AGNs emphasizes the need for efficient selection methods when conducting systematic searches for dual AGNs.}

\textcolor{black}{In our previous studies of dual AGNs, we preselected the late stage galaxy mergers (pair separations $<$ 10 kpc) based upon their mid-infrared \textit{WISE} colors \citep{satyapal2017,pfeifle2019}, selecting systems that exhibit a WISE $W1[3.4\mu\rm{m}]-\textit{W}2[4.6\mu\rm{m}]>0.5$ color, a color cut that has been demonstrated in simulations to be the most effective at identifying AGNs in late-stage mergers \citep{blecha2018}. We obtained follow-up high spatial resolution X-ray observations of these mergers with the \textit{Chandra X-ray Observatory} (\chandra{}) and longslit near-infrared spectroscopic observations with the Large Binocular Telescope (\textit{LBT}) to look for dual nuclear X-ray sources and high ionization coronal emission lines. Not only did a majority of the mergers host dual AGNs or dual AGN candidates \citep{ellison2017,satyapal2017,pfeifle2019}, but one system exhibited three nuclear X-ray sources and --- upon closer inspection --- was realized as a clear case of a triple merger: SDSS J084905.51+111447.2 (henceforth SDSS J0849+1114).\footnote{\textcolor{black}{In \citet{pfeifle2019}, two systems were reported as having triple nuclear X-ray sources: SDSS J0849+1114 and SDSS J1306+0735. While  the morphology of SDSS J0849+1114 is easily discerned, it is still unclear --- despite its three X-ray sources --- if SDSS J1306+0735 is a triple merger; a separate investigation is needed to better characterize this latter system.}} This study demonstrated the effectiveness of using mid-infrared colors as a pre-selection strategy for finding AGNs in advanced mergers.}

Dual AGNs in advanced mergers represent the most observationally accessible progenitors of the SMBH binary phase, which likely produces the main source of gravitational waves (GWs) \citep{wyithe2003,sesana2004,mingarelli2017,kelley2017b} detectable by Pulsar Timing Array campaigns \citep{verbiest2016} and future spaced-based observations  from the \textit{Laser Interferometer Space Antenna} (\textit{LISA}) \citep{amaro-seoane2017}. However, SMBH binaries in realistic astrophysical environments can stall on parsec-scale orbits, resulting in merger timescales that exceed the age of the Universe. An intruding third SMBH has been shown to significantly shorten coalescence time scales \citep{ryu2018} and can also result in slingshot ejections of one of the black holes from the host, resulting in either ejected or wandering SMBHs (e.g. \citealp{hoffman2007,bonetti2018a,bonetti2019}). Triple mergers are therefore of particular interest, because recent cosmological simulations predict 16\% of binary SMBHs resulting from major mergers will undergo an interaction with a third SMBH prior to coalescing \citep{kelley2017b}. \textcolor{black}{Triple AGN-hosting mergers are exceedingly rare, with only a handful of triple AGN candidates in mergers  reported in the literature \citep{liu2011,koss2012,kalfountzou2017} so far.}\footnote{\textcolor{black}{\citet{schawinski2012} reported the detection of a triple AGN candidate, but --- as stated by the authors --- given its high redshift, it is unclear if this system resides in a clumpy galaxy where the black holes are forming in situ, or if it is a merging system.}}$^,$\footnote{\textcolor{black}{\citet{deane2014} reported the detection of a triple AGN in the system SDSS J1502+1115 based on 1.7 GHz and 5 GHz observations from VLBI, however higher resolution follow-up observations presented in \citet{wrobel2014} suggest that this system more likely hosts a dual AGN where one of the AGNs exhibits radio hotspots.}}

\begin{figure}[t!]
 \centering
\includegraphics[width=0.48\textwidth]{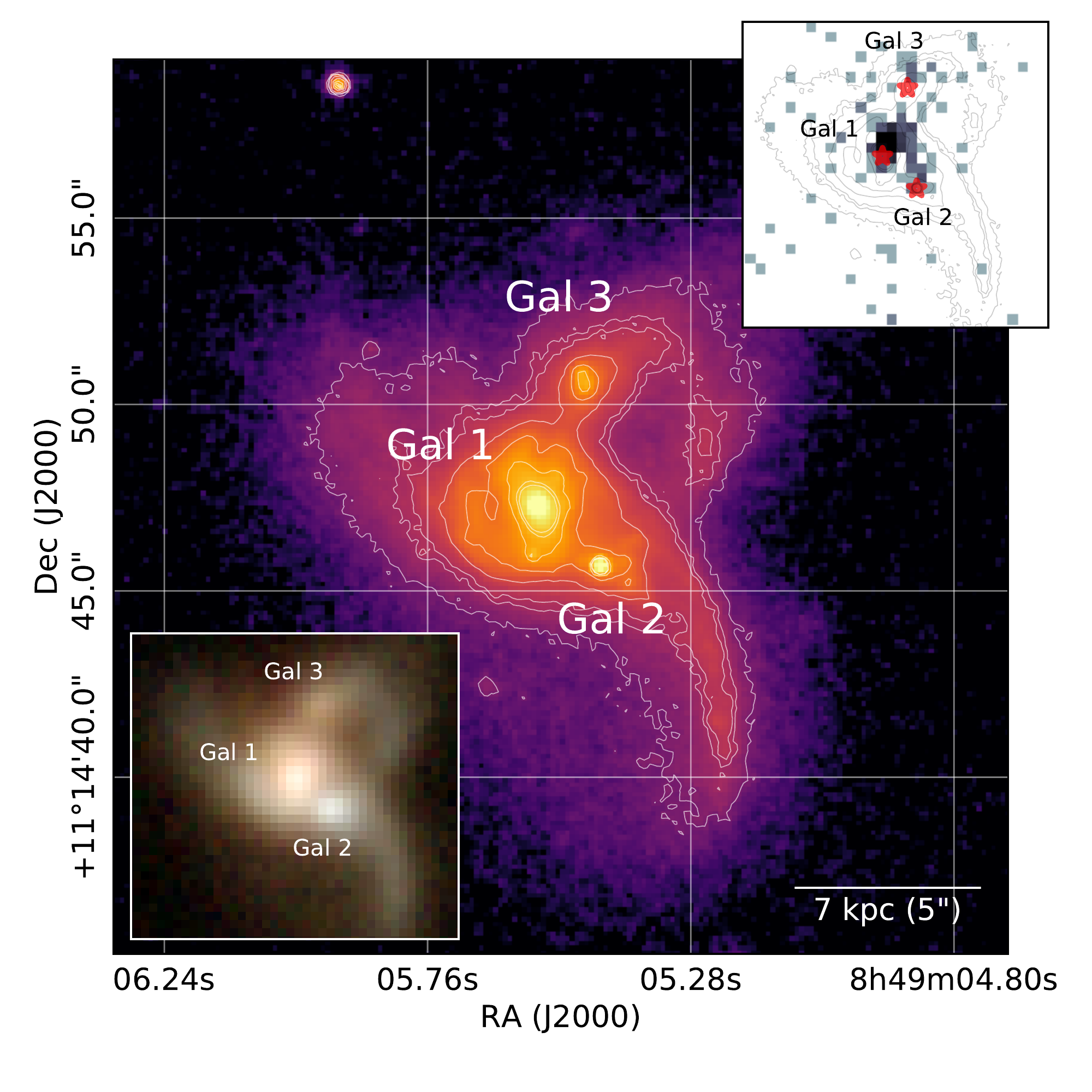}
 \caption{This figure shows the archival \textit{HST} WFC3 F105W image in the main figure and the combined \chandra{} $0.3-8$ keV X-ray image in the top right corner, with \textit{HST} contours overlaid on each. The bottom left corner displays the lower resolution \textit{SDSS} tricolor image. \textcolor{black}{We mark the approximate position of each of the three nuclei on the \chandra{} image with red stars.}
 }
\label{fig:hst_xrays}
\end{figure}

\textcolor{black}{In this work we present a suite of multiwavelength observations providing convincing evidence for a triple AGN in the merging system, SDSS J0849+1114.} SDSS J0849+1114 is a late stage  merger at a distance of $\sim$350 Mpc  ($\rm{z}=0.077$) comprising three interacting galaxies, with nuclear pair separations of 2.3\arcsec{} (3.4 kpc, $\rm{Galaxy\ 1}-\rm{Galaxy\ 2}$), 3.6\arcsec{} (5.3 kpc, $\rm{Galaxy\ 1}-\rm{Galaxy\ 3}$), and 5.0\arcsec{} (7.3 kpc, $\rm{Galaxy\ 2}-\rm{Galaxy\ 3}$) based upon archival \textcolor{black}{(PI: X. Liu)} \textit{Hubble Space Telescope} (\textit{HST}) imaging data (Figure~\ref{fig:hst_xrays}). Morphologically, the system exhibits strong tidal features indicative of an advanced merger. With an integrated 8-1000$\mu$m infrared luminosity of $\rm{log}(\rm{L}_{\rm{IR}}/\rm{L}_{\odot})=11.43\pm0.03$, this merger falls within the class of Luminous Infrared Galaxies (LIRGs). Based on the all-sky {\it Wide-field Infrared Survey Explorer} (\textit{WISE}), it displays a strong mid-infrared continuum with colors ($W1[3.4\mu\rm{m}]-\textit{W}2[4.6\mu\rm{m}]=1.69$) often associated with powerful AGN \citep{stern2012}, and was therefore included in our previous dual AGN sample. \textcolor{black}{As reported in \cite{pfeifle2019}, SDSS J0849+1114 exhibited three nuclear X-ray sources with luminosities in excess of that expected by star formation contributions, and a high ionization [SiVI] coronal  emission line was identified in the near-infrared spectrum of Galaxy 1. Furthermore, the system exhibited signs of high absorption along the line of sight based on its X-ray spectral properties.} 

\textcolor{black}{Interestingly, this object was also selected independently in a catalog of optically selected AGN pairs in \citet{liu2011}, and was identified as a triple AGN candidate. We combine the archival \chandra{} data from this program with our own in this analysis.}

\textcolor{black}{In this manuscript, we present new \nustar{} hard X-ray observations and X-ray spectral fitting results, new LBT optical spectroscopic observations along with archival SDSS optical spectra, we refine the source statistics using the combined \chandra{} datasets, and we show for the first time the near-IR spectra for all three galaxy nuclei. The manuscript is laid out as follows: In Section 2 we provide details on the observations and data reduction steps. In Section 3 we discuss the results obtained from the X-ray, near-IR, and optical spectroscopic observations. In Section 4 we explore alternative scenarios for the observed data and discuss this target in the broader context of our mid-infrared selection technique. In Section 5 we summarize our conclusions.} Throughout this paper we adopt $H_0=70$ km s$^{-1}$ Mpc$^{-1}$, $\Omega_M=0.3$, and $\Omega_\Lambda=0.7$.

\section{Observations and Data Reduction}
\textcolor{black}{During our original dual AGN campaign, we obtained \chandra{} X-ray observations and near-infrared spectroscopy from the LBT of SDSS J0849+1114. As part of a follow-up effort we also obtained hard X-ray \nustar{} observations and longslit optical spectroscopic observations from the LBT.\footnote{\textcolor{black}{For our LBT spectroscopic observations, we were careful to obtain observations in optimal seeing conditions ($<1\arcsec{}$) and extracted the spectra in apertures small enough ($1\arcsec{}$, see Section 2.4) to ensure measurements are restricted to each respective nucleus.}} During this analysis, we also retrieved the archival \textit{HST} and \chandra{} observations. The \chandra{} data were reduced using the same procedure outlined in-depth in \citet{pfeifle2019}.} Here we describe all other observations.

\subsection{\rm{NuSTAR} \textit{Imaging Observations}}

The 31ks \nustar{} observation was taken on 1 March 2017 (PI: Satyapal) and processed using the \nustar{} Data Analysis Software \textsc{(NuSTARDAS)}. To obtain level 2 data products, we ran \textsc{nupipeline} version 0.4.6 with the latest \textsc{caldb} and clock correction update files. Due to the moderate angular resolution of \nustar{} (FWHM 18\arcsec{}), we used a 30\arcsec{} radius aperture for source extraction and a 30\arcsec{} aperture for background extraction in the immediate vicinity of the source region. We then created the level three data products using \textsc{nuproducts}. The spectra generated for \textsc{fpma} and \textsc{fpmb} were then grouped by 1 count per bin, due to the low number of counts in the spectra, using \textsc{grppha}. 

\begin{figure*}[!ht]
    \centering
    \subfloat{\includegraphics[width=1.0\linewidth]{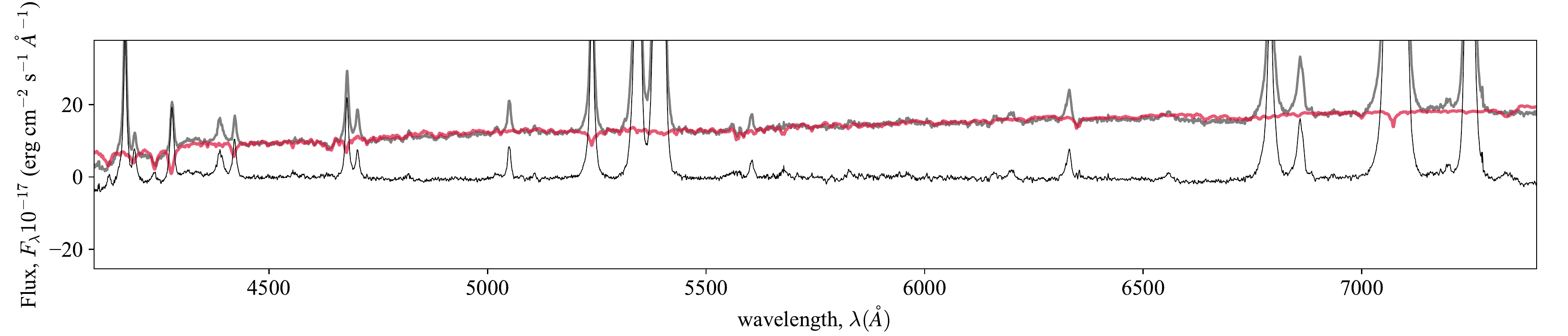}}\\ \vspace{-3mm}
    \subfloat{\includegraphics[width=1.0\linewidth]{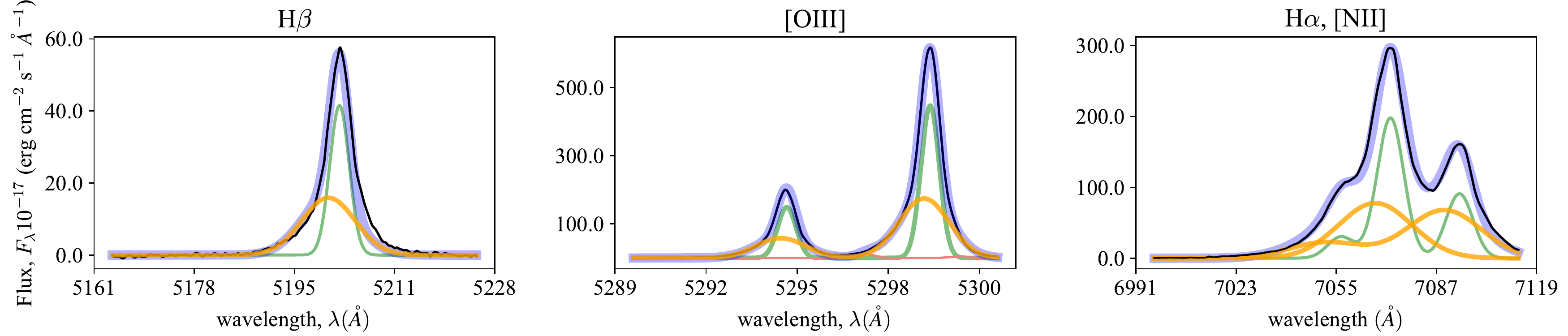}}\\ \vspace{-3mm}
    \subfloat{\includegraphics[width=1.0\linewidth]{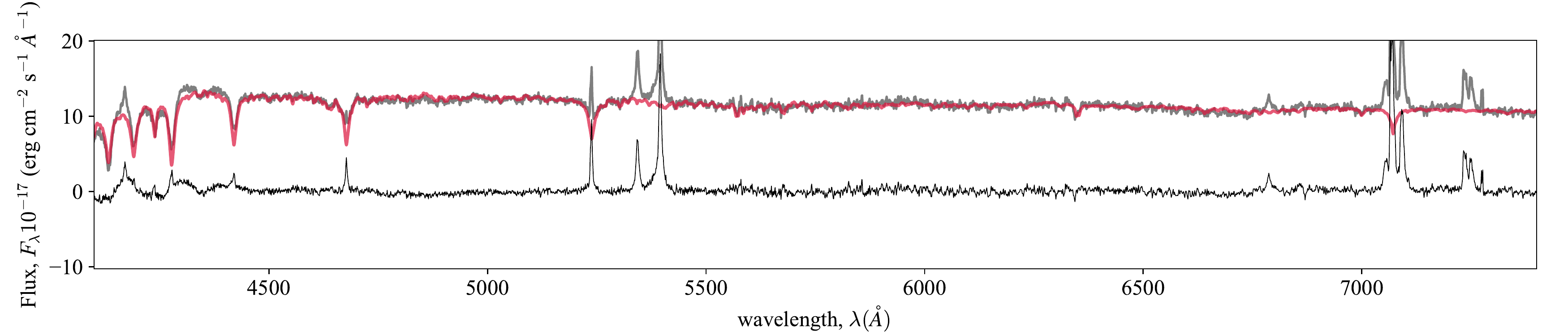}}\\ \vspace{-3mm}
    \subfloat{\includegraphics[width=1.0\linewidth]{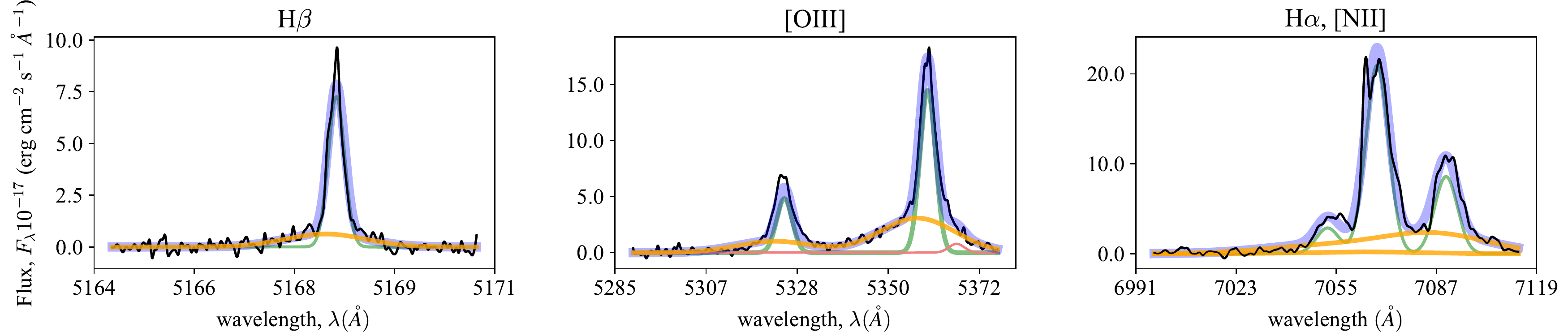}}\\ 
\caption{\textcolor{black}{The optical spectra of the three nuclei within SDSS J0849+1114, where rows one (Galaxy 1), three (Galaxy 2), and five (Galaxy 3) show the full \textcolor{black}{(observed frame)} optical spectrum of each nucleus, while rows two, four, and six show specific examples of emission line decompositions for Galaxy 1, 2, and 3. The spectra for Galaxy 1 and 2 were obtained with the MODS on LBT while the archival spectrum for Galaxy 3 was obtained from \textit{SDSS}. See the Appendix for a figure illustrating the full dynamic range of these spectra; here we show \textcolor{black}{the spectral region used for pPXF analysis} to aid the reader in examining the stellar population subtraction.} The spectra were fit using the procedure outlined in Section 2.5; the light gray spectrum in rows 1, 3, and 5 shows the full observed spectrum, the red line illustrates the spectrum of the best-fitting stellar population, and the black spectrum of each panel shows the resulting spectrum after subtracting the stellar population model. \textcolor{black}{The sets of panels illustrating the emission line decompositions are color-coded, where the orange shading represents the broad line component, the green shading represents the narrow line component, and the blue shading represents the combined (full) emission line. The solid black line in each decomposition panel shows the observed emission line from the spectrum after the stellar continuum subtraction.} \textcolor{black}{The thin red line in each [OIII]$\lambda5007$ decomposition panel shows the contaminating Fe I $\lambda$4985, $\lambda$4986, $\lambda$5016 emission.}}
\label{fig:opticallinefits}
\end{figure*}

\begin{figure*}[ht!]
    \ContinuedFloat
    \centering
    \subfloat{\includegraphics[width=1.0\linewidth]{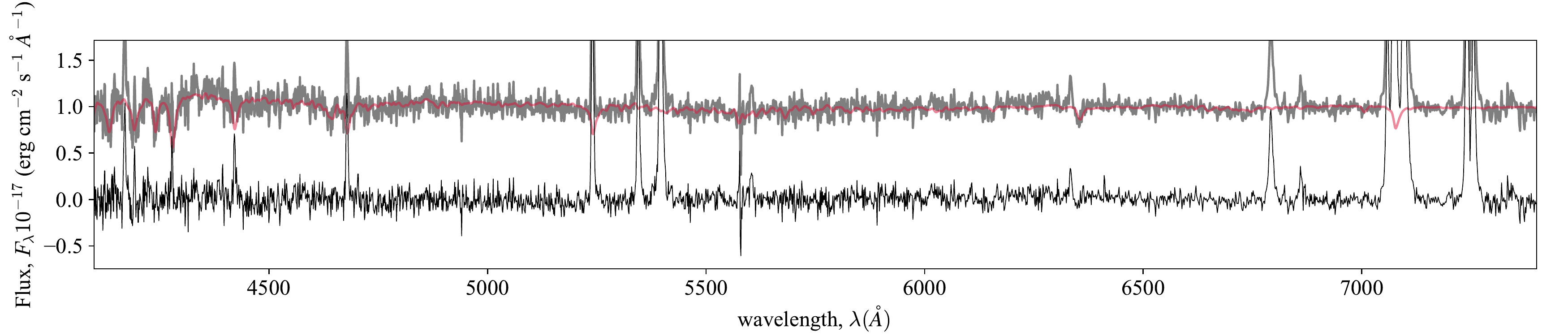}}\\
    \subfloat{\includegraphics[width=1.0\linewidth]{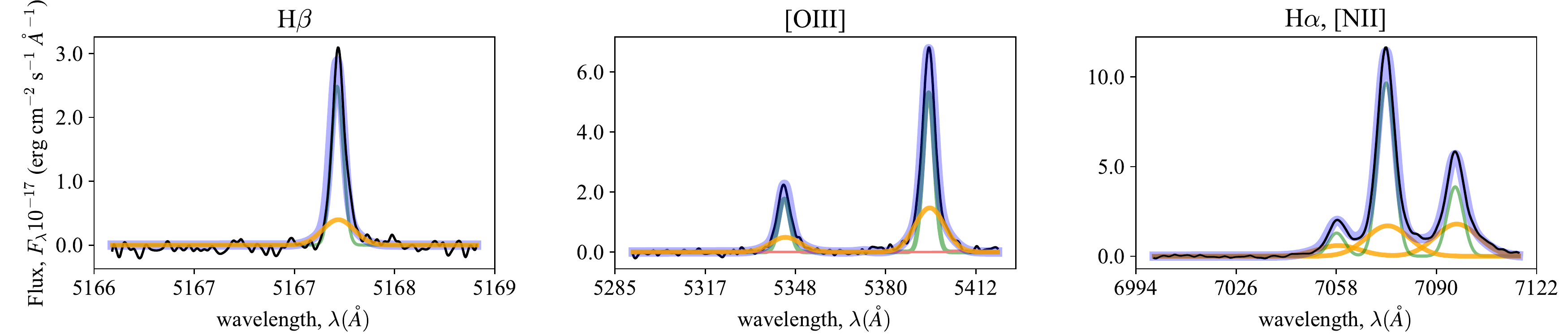}}\\
\caption{\textit{Continued}}
\end{figure*}

\subsection{\chandra{} and \nustar{} Spectral Analysis}

All simulations and X-ray spectral fitting for the \chandra{} and \nustar{} observations were conducted using the \textsc{xspec} \citep{arnaud1996} X-ray spectral fitting package version 12.9.1 with Cash statistics (C-Stat; \citealp{cash1979}). We employed a physically motivated model, \textsc{borus} \citep{balokovic2018}, for reprocessed emission along with model components for photoelectric absorption and Compton scattering. This model has 4 free parameters and is represented by: \textsc{constant$\times$ phabs(\textsc{borus}02 + zphabs $\times$ cabs $\times$ cutoffpl + constant $\times$ cutoffpl)}.

Due to the difference in the angular resolution limits of \chandra{} (PSF FWHM 0.5\arcsec{}) and \nustar{}, any hard X-ray photons from the three sources would be confused within the \nustar{} extraction aperture. Before any simultaneous fitting of the observations was performed we first attempted to deconvolve these signals. We began by simulating the $3-24$ keV spectra of the three \chandra{} sources using the \textsc{borus} model described above (with $\Gamma=1.9$), in each case adjusting the normalization such that the model returned the same $2-10$ keV fluxes. We simulated the spectra for a range of column densities (22.0 $\leq$ log($\nh{}/\rm{cm}^{-2}$) $\leq$ 25.0 in increments of log($\nh{}/\rm{cm}^{-2}\rm{)}=0.5$) and then derived the associated $3-24$ keV fluxes. We then fit the \nustar{} spectra with the same \textsc{borus} model; based upon the derived \nustar{} flux and the simulated fluxes of the three \chandra{} sources, it appeared that only Galaxy 1 contributes appreciably to the observed \nustar{} flux. We therefore fit the \nustar{} data simultaneously with the \chandra{} spectrum of the Galaxy 1 source, only. \textcolor{black}{It is not possible to constrain whether the two remaining AGN are Compton-thick, and it is still possible that they contribute a non-negligible fraction of the \nustar{} flux.}

\subsection{\rm{LBT} \textit{Near-Infrared Spectroscopic Observations}}
Observations were obtained using LUCI-2 for Galaxy 1 and 2 on 30 January 2018 simultaneously (total of 1440 seconds) using a single slit position with a position angle (PA) of 49.4$^{\circ}$ and for Galaxy 3 on 03 January 2018 (total of 2400 seconds) using the same PA. All observations used a 1\arcsec{}.5 wide slit and the low-resolution G200 grating with HKspec filter. This provides a wavelength coverage of 1.46$\mu$m $<$ $\lambda$ $<$ 2.34$\mu$m and R $\simeq$ 570-920. The seeing was $<$ 1\arcsec{} for all observations. The data were reduced using customized IRAF scripts for flat-fielding, wavelength calibration, and rectification of the spectra.  Galaxies 1 and 2 were extracted in a 3-pixel (0\arcsec{}.75) wide slit and a 4.8 pixel (1\arcsec{}.2) wide slit for Galaxy 3. The spatial apertures were selected to be larger than the seeing to maximize signal-to-noise.  Telluric corrections and flux calibration was performed on each galaxy using Version 4.0 of Xtellcor \citep{vacca2003}.

\subsection{\rm{LBT} \textit{Optical Spectroscopic Observations}}

Binocular \textcolor{black}{Multi-Object Double Spectrograph (MODS)} observations of Galaxy 1 and 2 were obtained simultaneously on 28 November 2018 using a single slit position with PA = 230$^{\circ}$. MODS-1 was configured with the dual grating (blue and red channels) and MODS-2 configured with the blue channel only. Observations were obtained with a 1\arcsec{}.2 wide slit and a total of 3600 seconds for the red channel and 7200 seconds combined for the two blue channels.  The average seeing during the observations was $<$ 1\arcsec{}. The data were reduced first using Version 2.04 of the modsCCDRed python scripts and then flat-fielded, wavelength calibrated, rectified, and flux-calibrated using customized IRAF scripts.  A 1D spectrum for each galaxy was extracted in a 1\arcsec{} wide spatial diameter based on the same criteria as the near-IR spectra.

\subsection{\rm{LBT} \textit{Optical Spectral Fitting to Determine Fluxes and Stellar Velocity Dispersion}}

In order to measure accurate line fluxes, it is necessary to account for stellar absorption, which primarily affects the Balmer lines. The Penalized Pixel-Fitting method (pPXF; \citealp{cappellari2017}) is an algorithm developed to extract stellar kinematics using a maximum penalized likelihood approach to match stellar spectral templates to absorption features in galaxy spectra. A large ($>150$) number of stellar templates of various spectral type were chosen from the Indo-US Library of Coud\'e Feed Stellar Spectra \citep{valdes2004} in order to decrease the possibility of template mismatch.  To ensure that the Balmer absorption is adequately constrained, we model the stellar population using the full spectrum (4000-7500\AA) with pPXF.  After subtracting the best-fitting stellar population model (red, see Figure~\ref{fig:opticallinefits})\footnote{\textcolor{black}{See the Appendix for a figure illustrating the full dynamic range of the optical spectra.}} from the continuum, the residual emission lines were fit using Gaussian models. Broadened wing components, which can be indicative of high-velocity gas outflows, are present in narrow lines of all three objects, and thus an additional Gaussian component is added  \textcolor{black}{(see Figure~\ref{fig:opticallinefits} for examples of emission line decompositions)}. The presence of Fe emission can also affect measurements of the stellar population and [\ion{O}{3}]$\lambda\lambda$4959,5007 doublet, and therefore we include the \ion{Fe}{2} template from \citet{veroncetty2004} and \ion{Fe}{1} lines from the NIST Atomic Spectra Database in our fits. Fitting was performed using a custom maximum-likelihood routine implemented in Python, which uses the affine-invariant MCMC ensemble sampler \textit{emcee} \citep{foremanmackey2013}.  Line fluxes are determined from the best-fit Gaussian models of the emission lines.  Stellar-velocity dispersions were fit using a similar method described, except we fit the region from 4400-5800\AA, which includes the Mg Ib region used to estimate stellar velocity dispersion. \textcolor{black}{The [OIII] doublet is best fit by a combination of two Gaussian components. The line widths from the [OIII] doublet fit are then used to constrain the widths of the rest of the emission lines \citep{rodriguez2013}, so careful modeling of their profiles is necessary. The fitting routine can actually overestimate the flux and width of the broad component of the [OIII]$\lambda$5007 emission line due to contaminating flux from the Fe I $\lambda$4985, $\lambda$4986, $\lambda$5016 emission lines. To address this contamination, we model the Fe lines with independent amplitudes and with widths equal to that of the narrow component of [OIII] \citep{manzanoking2019}.}


\section{Results}
\subsection{\rm{Chandra}/\rm{ACIS-S} \textit{Imaging Results}}

We combined our 2016 observation with an archival 2013 observation (PI: X. Liu), and assessed source significance using the binomial no-source probability statistic (P$_B$) \citep{lansbury2014} which is suitable for the low-count regime; we point the reader to \citet{pfeifle2019} for further discussion. \textcolor{black}{The \chandra{} source statistics are reported in Table~\ref{table:xraystats}.} The Galaxy 1, 2, and 3 sources were detected with a statistical significance of 13.8$\sigma$ [log(P$_B\rm{)} = -371.6$], 3.5$\sigma$ [log(P$_B\rm{)}=-17.3$], and 2.9$\sigma$ [log(P$_B\rm{)}=-12.0$], and based upon the derived P$_B$ values it is highly unlikely that these sources are the result of spurious background activity.\footnote{\textcolor{black}{These significance values differ slightly from those reported in \citet{pfeifle2019} due to the fact that here we quote the statistical significance derived from the combined \chandra{} statistics, whereas previously the statistical significance was derived from only the 2016 \chandra{} dataset.}} We used the \textsc{ciao} \textsc{modelflux} package to uniformly estimate the hard $2-10$ keV (rest frame) luminosities for the sources, assuming $\Gamma=1.9$ and correcting for only Galactic $\nh{}$ along the line of sight. Galaxies 1, 2 and 3 exhibit $2-10$ keV luminosities of $2.37\pm0.17\times10^{41}$ erg s$^{-1}$, $2.4^{+0.8}_{-0.7}\times10^{40}$ erg s$^{-1}$, $1.5^{+0.6}_{-0.5}\times10^{40}$ erg s$^{-1}$ respectively.

\begin{table*}
\caption{\chandra{} X-ray Source Statistics}
\begin{center}
\hspace{-2.2cm}
\begin{tabular}{cccccccccc}
\hline
\hline
\noalign{\smallskip}
Galaxy  & $\nh$ & $\alpha_\chi$ & $\delta_\chi$ & Counts  & Counts  & Counts  & HR & $\sigma$ & log$(P_{\rm{B}})$ \vspace{-3mm}\\
\noalign{\smallskip}
&  &  &  &  &  &  &  & & \vspace{-3mm}\\
\noalign{\smallskip}
& ($10^{20}$ cm$^{-2}$) &  &  &  $0.3-8$ keV &  $0.3-2$ keV &  $2-8$ keV & & &\\
\noalign{\smallskip}
\hline
\noalign{\smallskip}
1 & 3.80 & 8\hr49\min05\sec.529 & +11\deg14\arcmin47\arcsec.876 & $196\pm14$ & $112\pm11$ & $84\pm9$ & -0.14 & 13.8$\sigma$ & -371.6 \\
2 & 3.80 & 8\hr49\min05\sec.381 & +11\deg14\arcmin45\arcsec.747 & $15^{+5}_{-4}$ & $13^{+4}_{-2}$ & $2^{+3}_{-1}$ & -0.76 & 3.5$\sigma$ & -17.3 \\
3 & 3.80 & 8\hr49\min05\sec.448 & +11\deg14\arcmin51\arcsec.646 & $13^{+5}_{-4}$ & $12^{+5}_{-4}$ & $1^{+2}_{-1}$ & -0.84 & 2.9$\sigma$ & -12.0 \\
\noalign{\smallskip}
\hline
\end{tabular}
\end{center}
\tablecomments{\textcolor{black}{ Column 1: X-ray source and galaxy nucleus designation. Column 2: Galactic hydrogen column density along the line of sight. Column 3-4: Right ascension (RA) and declination (DEC) of the source. Column 5-7: X-ray counts detected in the full, soft, and hard X-ray bands for the combined 2013 and 2016 \chandra{} observations. Note, these are not background-subtracted counts, but formal statistical significances of the sources are derived from background subtracted statistics. Values are rounded to closest integer. Column 8: Hardness ratio, given by the formula (H-S)/(H+S), where H and S refer to the \chandra{} hard and soft energy bands. Column 9: Formal source statistical significance, derived from background subtracted source statistics. Column 10: Logarithm of the binomial source probability, which is a metric for source statistical significance more appropriate for the low-count regime.} \textcolor{black}{Note: Sources which meet a threshold of $\rm{log}(P_{\rm{B}})<-2.7$ are considered to be true sources, and it is highly unlikely these are the result of spurious background activity \citep{lansbury2014}; we used this statistical metric for our dual sample in both \citet{satyapal2017} and \citet{pfeifle2019}.}}
\label{table:xraystats}
\end{table*}

\begin{table*}
\caption{\nustar{} X-ray Source Statistics}
\begin{center}
\hspace{-2.2cm}
\begin{tabular}{ccccccc}
\hline
\hline
\noalign{\smallskip}
Camera & $\nh$ & $\alpha_\chi$ & $\delta_\chi$ & Counts  & Counts  & Counts \vspace{-3mm}\\
\noalign{\smallskip}
 & & & & & & \vspace{-3mm}\\
\noalign{\smallskip}
 & & & & $3-78$ keV & $3-8$ keV  &  $8-24$ keV  \\
\noalign{\smallskip}
\hline
\noalign{\smallskip}
FPMA & 3.80 & 8\hr49\min05\sec.510 & +11\deg14\arcmin47\arcsec.260 & $141\pm12$ & $20\pm4$  & $63\pm8$  \\
FPMB & 3.80 & 8\hr49\min05\sec.510 & +11\deg14\arcmin47\arcsec.260 & $149\pm12$ & $49\pm7$ & $59\pm8$ \\
\noalign{\smallskip}
\hline
\end{tabular}
\end{center}
\tablecomments{\textcolor{black}{Column 1: \nustar{} camera. Column 2: Galactic hydrogen column density along the line of sight. Column 3-4: RA and DEC coordinates of the source apertures. Column 5: X-ray counts in the $3-78$ keV \nustar{} band. Column 6: X-ray counts in the $3-8$ keV \nustar{} band. Column 7: X-ray counts in the $3-24$ keV \nustar{} band.}}
\label{table:nustarstats}
\end{table*}

\subsection{\rm{Chandra}/\rm{ACIS-S} \textit{and} \rm{NuSTAR} \textit{Spectral Analysis Results}}
\begin{figure}
  \centering
    \includegraphics[width=0.5\textwidth]{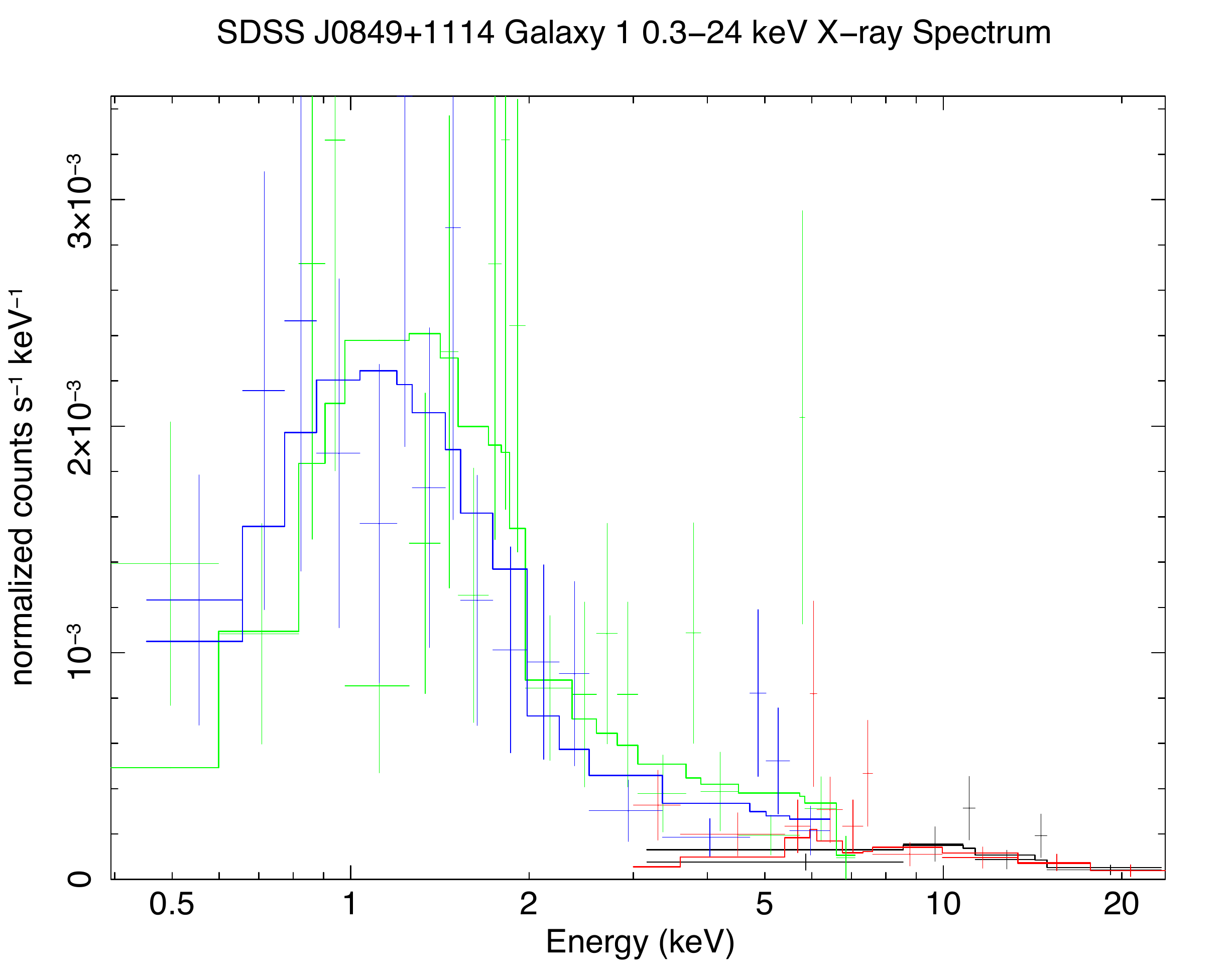}
    \caption{\textcolor{black}{This diagram depicts the \nustar{} 3-24 keV (Black: FPMA, Red: FPMB; PI: Satyapal) and two \chandra{} 0.3-8 keV spectra (Green: PI, Satyapal; Blue: PI, Liu) along with the best fitting model to the data. Fitting with the Borus reprocessed emission model, along with components for Compton-scattering and reflected emission, reveals that the Galaxy 1 source is a heavily obscured AGN.}}
    \label{fig:xrayfitting}
\end{figure}

Simultaneously fitting the \chandra{} and \nustar{} data for Galaxy 1 \textcolor{black}{(see Figure~\ref{fig:xrayfitting})}, we find an absorbing column of log($\nh{}$/cm$^{-2}$) $=23.9^{+0.2}_{-0.2}$, \textcolor{black}{photon index} $\Gamma=1.4^{+0.3}_{\rm{NC}}$, and a scattered fraction $f_S=11.7^{+22.1}_{-6.4}$\%.\footnote{NC = Not constrained} Correcting for the intrinsic absorption, we find an unabsorbed $2-10$ keV luminosity of $4.6^{+0.6}_{-0.6}\times10^{42}$ erg s$^{-1}$. \textcolor{black}{We also report the \nustar{} source statistics in Table~\ref{table:nustarstats}. To measure the equivalent width of the Fe K$\alpha$ line, we repeated the fitting process with the phenomenological model from \citet{pfeifle2019}, which includes a Gaussian emission line component to account for the iron line fluorescent emission. Due to the S/N of our X-ray data, we conservatively quote here only an upper limit for the Fe K$\alpha$ emission. We find an equivalent width upper limit of $\sim0.6$ keV for the Fe K$\alpha$ line, which is consistent with the high absorbing column along the line of sight \citep{brightman2011}. }

\textcolor{black}{While we do find different values for the spectral parameters here than those reported in \citet{pfeifle2019}, this is not entirely surprising; in this manuscript we are fitting both sets of \chandra{} and \nustar{} observations simultaneously, and it is well known that \nustar{} observations aid significantly in constraining the values for spectral parameters such as the photon index, $\nh{}$, and the equivalent width of the Fe K$\alpha$ line \citep{marchesi2018}, in some cases leading to refined values significantly different than those found with data in the $2-10$ keV energy range.}

\textcolor{black}{As we discussed above in Section 2.2, it is not possible to constrain whether the two remaining AGNs are Compton-thick, and it is still possible they contribute appreciably to the observed hard X-ray flux since they would fall within the \nustar{} extraction aperture.} In any case, this does not affect our X-ray spectral fitting results; the results are consistent with the picture of mergers hosting obscured AGNs, as found by previous studies \citep{kocevski2015,ricci2017mnras,ricci2016,lanzuisi2018,goulding2018,donley2018,koss2018}.

\subsection{LBT Spectroscopic Results}

\begin{figure*}[!ht]
    \centering
    \subfloat{\includegraphics[width=0.34\linewidth]{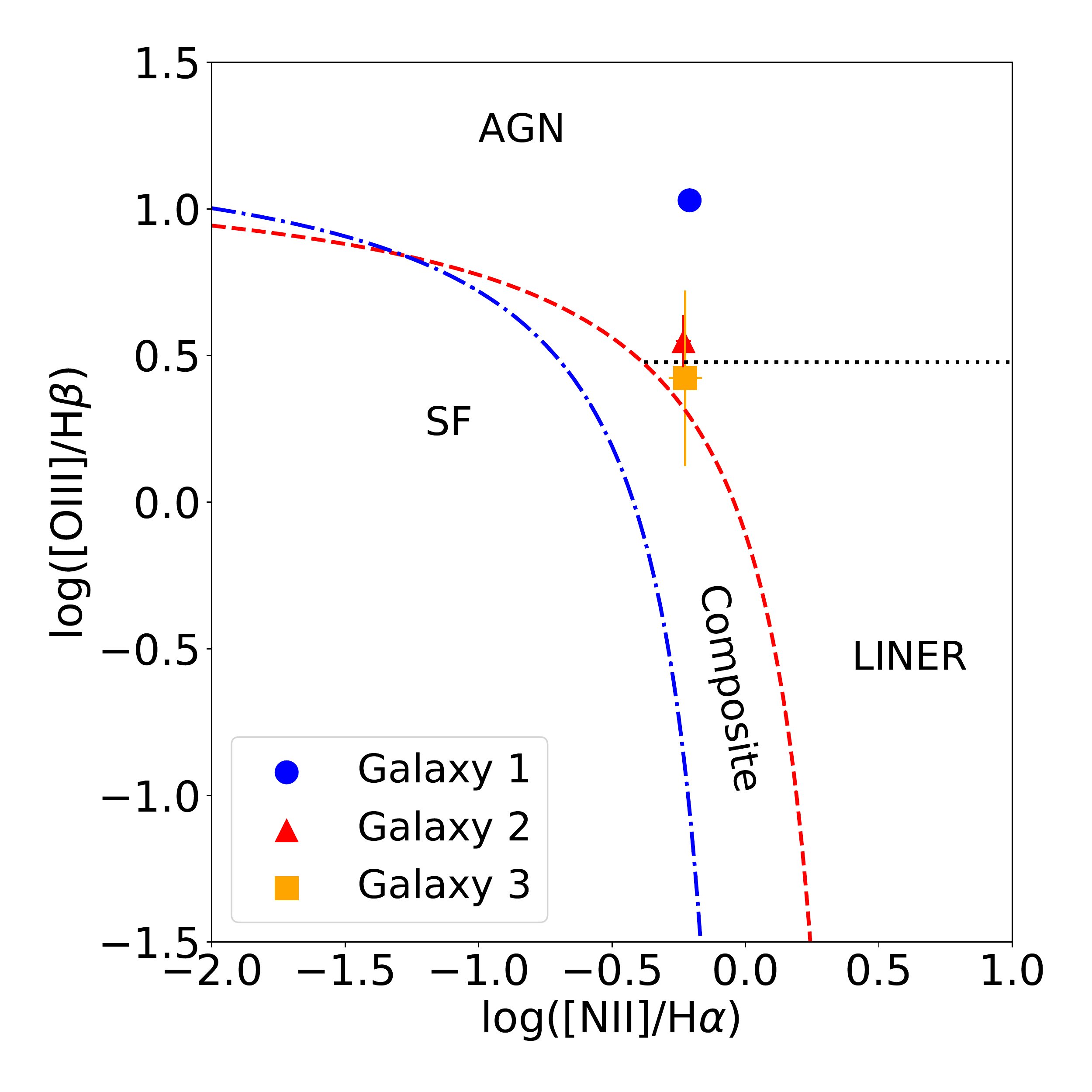} \hspace{-3mm}} 
    \subfloat{\includegraphics[width=0.34\linewidth]{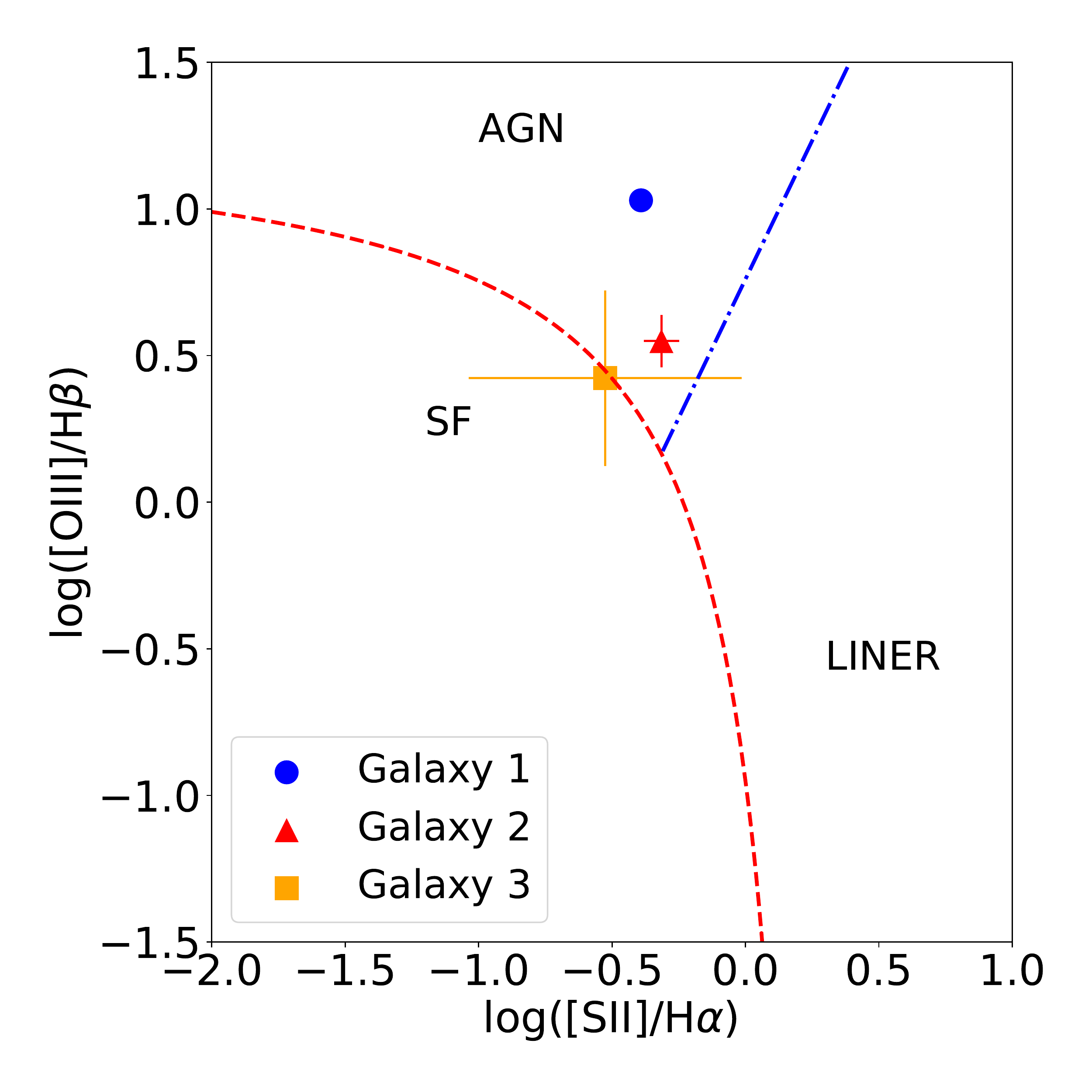} \hspace{-3mm}} 
    \subfloat{\includegraphics[width=0.34\linewidth]{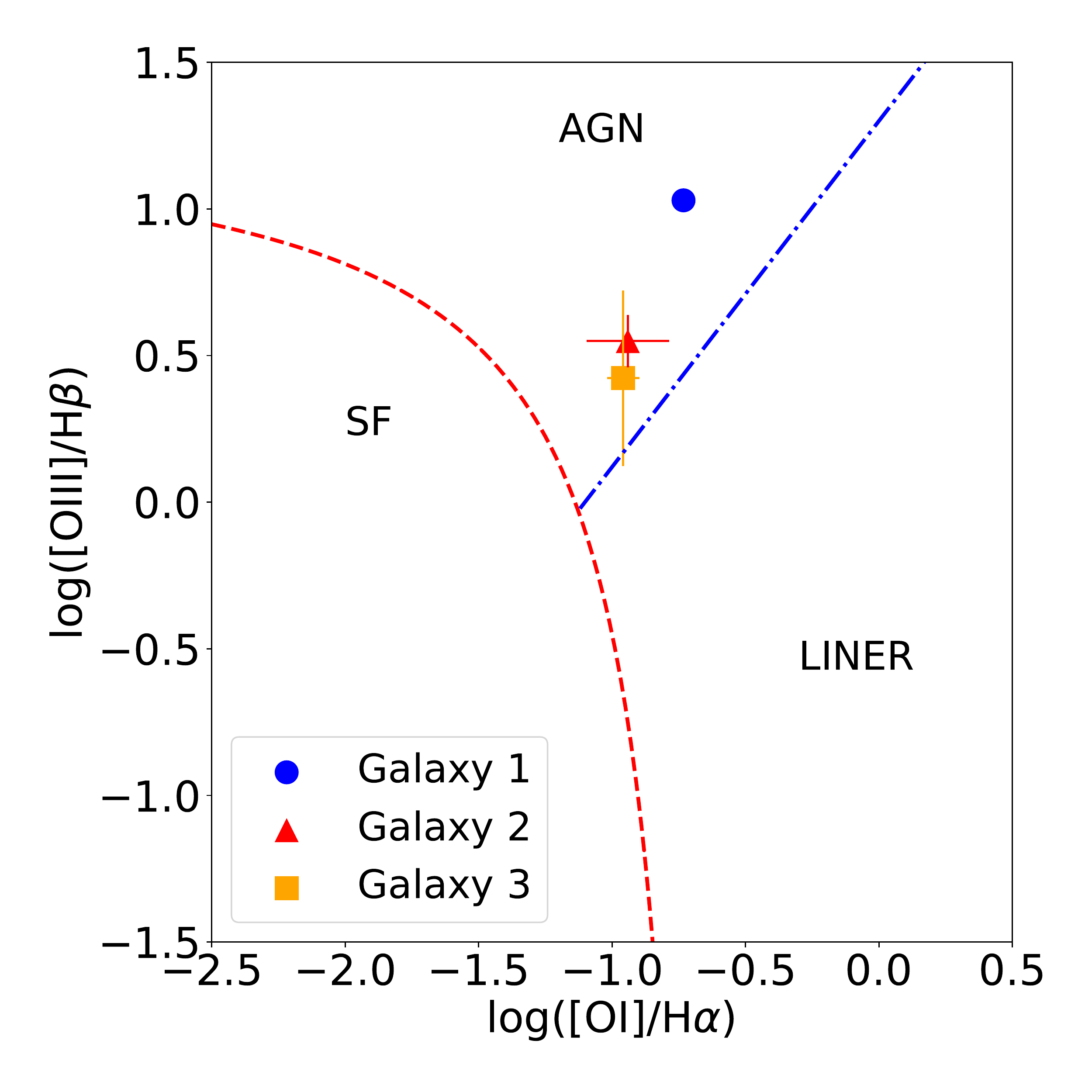}}\\ \vspace{-2mm}
\caption{Three BPT optical spectroscopic line ratio diagrams, based on the [OIII]$\lambda$5006/H$\beta$ to [NII]/H$\alpha$, [SII]$\lambda\lambda$6717,6733/H$\alpha$ doublet, and [OI]$\lambda$6302/H$\alpha$ emission line ratios. (Left): The red (dashed), blue (dash-dotted), and black (dotted) lines represent the \citet{kewley2001}, \citet{kauffmann2003}, and \citet{ho1997} demarcations, respectively, which separate AGN, star-forming, and LINER populations. (Center and right): The red (dotted) and blue (dashed) lines represent the \citet{kewley2001} and \citet{kewley2006} demarcations, which separate HII regions, Seyferts, and LINERs. All three nuclei display AGN-like line ratios. The [SII] and [OI] ratios are inconsistent with shock-driven photoionization \citep{dopita1995}, which would typically produce line ratios in the lower right (LINER) corners of the middle and right diagrams. \textcolor{black}{The logarithmic emission line ratios plotted above are listed in Table~\ref{table:lbtstats}.\\}}
\label{fig:BPTs}
\end{figure*}

 To determine the nature of the photoionizing sources observed in the LBT optical spectroscopy, we employ "Baldwin, Phillips, \& Terlevich" (BPT) optical spectroscopic diagnostic diagrams, shown in Figure~\ref{fig:BPTs}. The optical emission in each nucleus is consistent with AGN photoionization, complementing the X-ray and near-IR observations discussed above, and is consistent with the presence of a triple AGN in this merger. We list the logarithmic emission line ratios in Table~\ref{table:lbtstats}.

\textcolor{black}{Following the procedure outlined in Section 2.5, we derived stellar and gas velocity dispersions for each nucleus, which we report in Table~\ref{table:lbtstats}. Gas velocity dispersions were measured from the narrow component of the [OIII]$\lambda5007$ emission line while the stellar velocity dispersions were measured by fitting the Mg Ib region in the spectral range $4400-5800$\AA. We find very similar values for the stellar and gas velocity dispersions within each respective nucleus \textcolor{black}{(see Columns $6-7$ in Table~\ref{table:lbtstats}), but these values differ between the three nuclei}; these similarities  further suggest that each of the three regions are more-or-less dynamically independent --- that is, the gas and stellar components are well-coupled --- especially for Galaxies 1 and 3.\footnote{\textcolor{black}{We caution the reader that the interpretation of velocity dispersions is highly uncertain in merging systems \citep{stickley2014}. We list these dispersions here simply to support the scenario of three kinematically distinct nuclei.}}}

\textcolor{black}{We observe strong blue-wing components [OIII]$\lambda$5007 emission lines in Galaxy 1 and 2 with FWHMs on the order of $\sim1300-1600$ km s$^{-1}$, which is suggestive of strong galactic-scale outflows \citep{heckman1981,nelson1996,woo2016}. Galaxy 3 shows a slightly redshifted wing in the [OIII]$\lambda$5007 emission, with a FWHM on the order of $\sim700$ km s$^{-1}$. Many of the lower ionization emission lines show broad-wing components as well.}

\begin{figure}[t!]
 \centering
\includegraphics[width=0.47\textwidth]{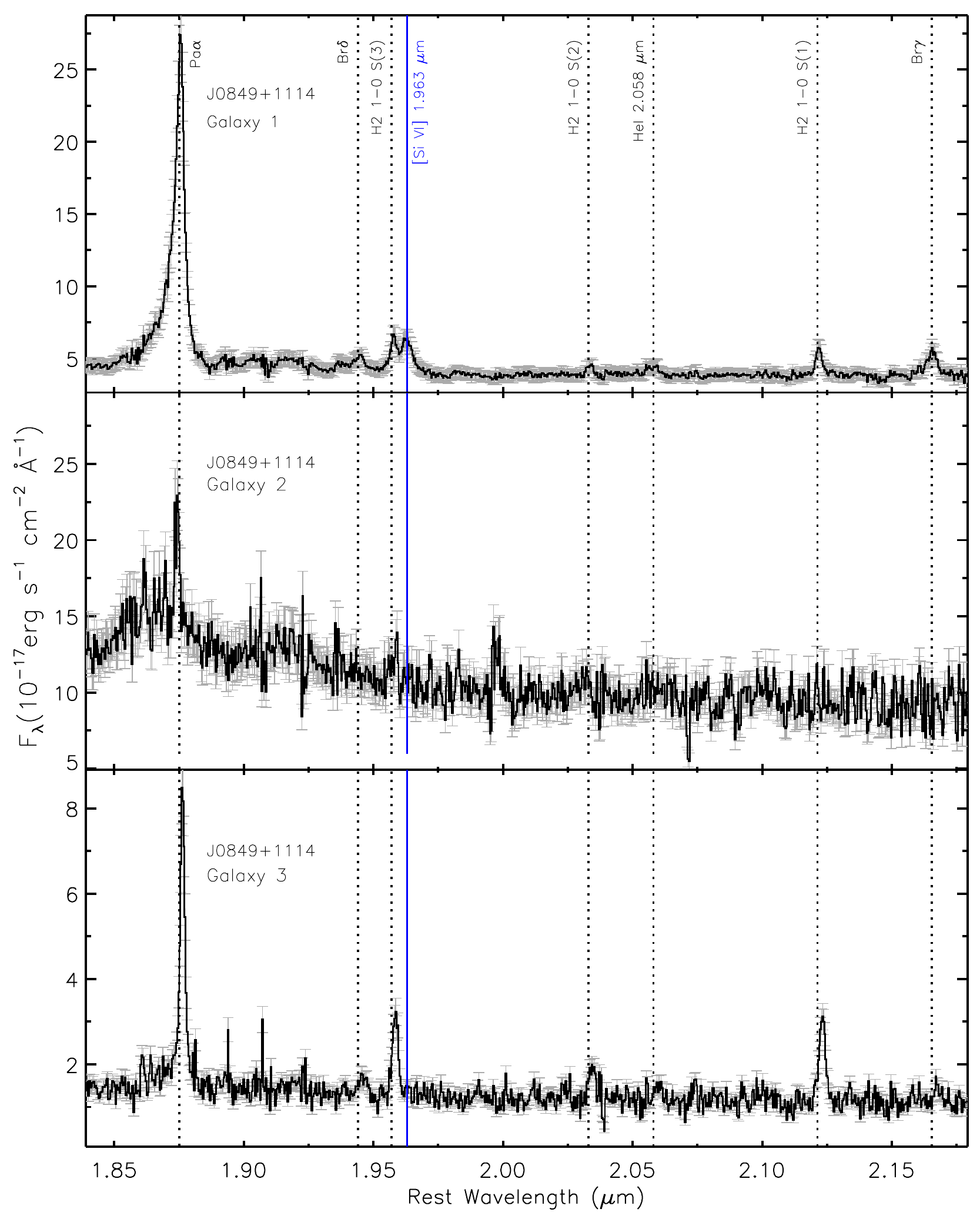}
 \caption{\textcolor{black}{These three panels show the near-IR K-band spectra for the three nuclei in SDSS J0849+1114 obtained from the LBT, where the top, middle, and bottom panels correspond to the spectra for Galaxy 1, 2 and 3. Dotted black lines indicate Hydrogen and Helium line emission, while the blue line indicates the wavelength corresponding to the [SiVI]1.967$\mu$m high ionization coronal emission line. One [SiVI]1.967$\mu$m coronal line is detected in Galaxy 1 ($\sigma$ = 2.3). The detection of a coronal line provides robust confirmation for the AGN in this nucleus.}}
\label{fig:nearIR_CLs}
\end{figure}

\begin{figure}[t!]
 \centering
\includegraphics[width=0.45\textwidth]{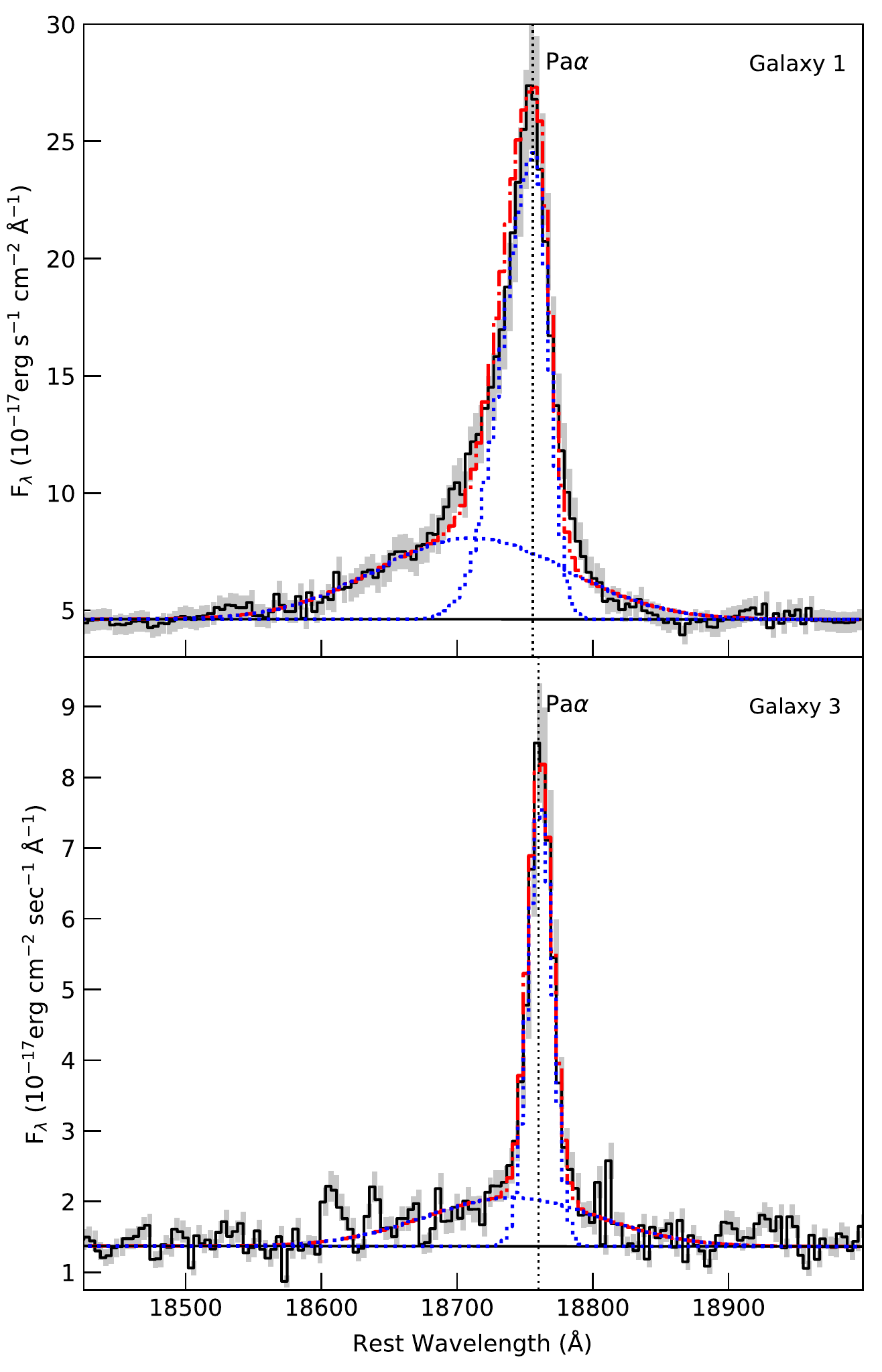}
    \caption{\textcolor{black}{These panels show profile fits for the Pa$\alpha$ near-infrared emission in the spectra of Galaxy 1 and 3. These broad components (FWHM $\gtrsim2400$ km s$^{-1}$) in the Pa$\alpha$ emission are significantly in excess of the blue-shifted components observed in the [OIII]$\lambda5007$ emission, consistent with a broad line region in these two galaxies. Note, the quality of the spectra in Galaxy 2 was insufficient for an emission line decomposition.}}
\label{fig:broadPa-alpha}
\end{figure}

\textcolor{black}{The LBT near-IR spectra revealed high excitation [SiVI] \citep{pfeifle2019} coronal line emission ($\rm{f}_\lambda=1.63\pm0.07\times10^{-15}\ \rm{erg}\ \rm{cm}^{-2}\ \rm{s}^{-1}$) in the nucleus of Galaxy 1, providing robust confirmation for the AGN in that nucleus (see Figure~\ref{fig:nearIR_CLs}). Moreover, the near-infrared spectra of Galaxy 1 and Galaxy 3 reveal broad components (FWHM $\gtrsim2400$ km s$^{-1}$) in the Pa$\alpha$ emission (see Figure~\ref{fig:broadPa-alpha}), significantly in excess of the blue-shifted components observed in the [OIII]$\lambda5007$ emission. These detections offer additional evidence for the existence of black hole accretion in these galaxies. We point out that this accretion activity is likely missed in the optical due to large obscuring columns along the line of sight. An exhaustive analysis of the possible outflows observed in the optical spectra is, however, beyond the scope of this paper, but will be featured in a future publication focusing on the environmental impacts of AGNs in mergers.}

\begin{table*}
\caption{LBT and SDSS Spectroscopic Results}
\begin{center}
\hspace{-2.2cm}
\begin{tabular}{ccccccc}
\hline
\hline
\noalign{\smallskip}
Galaxy & log([OIII]$\lambda$5006/H$\beta$) & log([NII]/H$\alpha$) & log([SII]$\lambda\lambda$6717,6733/H$\alpha$) & log([OI]$\lambda$6302/H$\alpha$) & $\sigma_*$ & $\sigma_{\rm{[OIII]}}$ \\
\noalign{\smallskip}
 & & & & & &\vspace{-5mm}\\
\noalign{\smallskip}
 & & & & & km s$^{-1}$ & km s$^{-1}$ \\
\noalign{\smallskip}
\hline
\noalign{\smallskip}
1 & $1.029\pm0.009$ & $-0.209\pm0.003$ & $-0.39\pm0.01$ & $-0.733\pm0.006$ & $180\pm9$ & $191\pm1$ \\
2 & $0.55\pm0.09$ & $-0.232\pm0.03$ & $-0.31\pm0.07$ & $-0.9\pm0.2$ & $149\pm11$ & $125\pm4$ \\
3 & $0.4\pm0.3$ & $-0.23\pm0.06$ & $-0.5\pm0.5$ & $-0.96\pm0.06$ & $91\pm26$ & $91\pm26$ \\
\noalign{\smallskip}
\hline
\end{tabular}
\end{center}
\tablecomments{\textcolor{black}{Column 1: Galaxy nucleus designation. Columns 2-5: Logarithmic optical emission line ratios for each nucleus, calculated using the derived optical emission line fluxes. See Section 2.5 for more details on how these line fluxes were derived from the optical spectra. Column 6: Stellar velocity dispersion. Column 7: Gas velocity dispersion. The large uncertainties in gas and velocity dispersions for Galaxy 3 are due to the lower signal-to-noise of SDSS spectra compared to LBT-obtained spectra.}}
\label{table:lbtstats}
\end{table*}

\section{Discussion and Implications}
While our multiwavelength observations point to each nucleus hosting an AGN, we considered alternative scenarios which might explain the observed characteristics of this system.

In order to rule out star formation as an alternative explanation for the observed X-ray emission, we calculated star formation rates (SFRs) and possible X-ray contributions from a population of X-ray binaries using the Pa$\alpha$ emission observed in the LBT near-IR spectra. We refer the reader to \citet{satyapal2017} and \citet{pfeifle2019} for a discussion of the SFR and XRB contribution calculations. \textcolor{black}{With this calculation, we estimate an upper limit for the star formation activity in the merger nuclei; in reality, since the nuclei all exhibit BPT AGN-like optical spectroscopic line ratios, some fraction of the Pa$\alpha$ emission is in fact attributable to the AGNs and therefore the true star formation rate, and hence stellar contribution to the X-rays, is much lower.}

We obtained SFRs of 13.16 M$_\odot$ yr$^{-1}$, 0.48 M$_\odot$ yr$^{-1}$, and 1.79 M$_\odot$ yr$^{-1}$ and predicted hard $2-10$ keV X-ray luminosities of $2.27\pm0.4\times10^{40}$ erg s$^{-1}$, $0.12\pm0.11\times10^{40}$ erg s$^{-1}$, and $0.43\pm0.20\times10^{40}$ erg s$^{-1}$ for Galaxy 1, 2, and 3, respectively. Contrasting these values with the observed X-ray emission (uncorrected for intrinsic absorption) in Galaxy 1, 2, and 3 (see Section 3.1), there is a clear order of magnitude disparity between the observed and predicted values, strongly suggestive of three AGNs. \textcolor{black}{Once again we note that in reality --- due to the expected contributions from the AGNs to the Pa$\alpha$ flux --- we are likely overestimating the X-ray contributions from SF. Moreover, we are underestimating the observed X-ray fluxes of the three sources as the quoted luminosities are not corrected for absorption}\textcolor{black}{; as indicated by our joint \nustar{}-\chandra{} spectral fitting results, any absorption corrections made to the observed X-ray luminosities are expected to be large.}

\begin{figure}[t!]
 \centering
\includegraphics[width=0.48\textwidth]{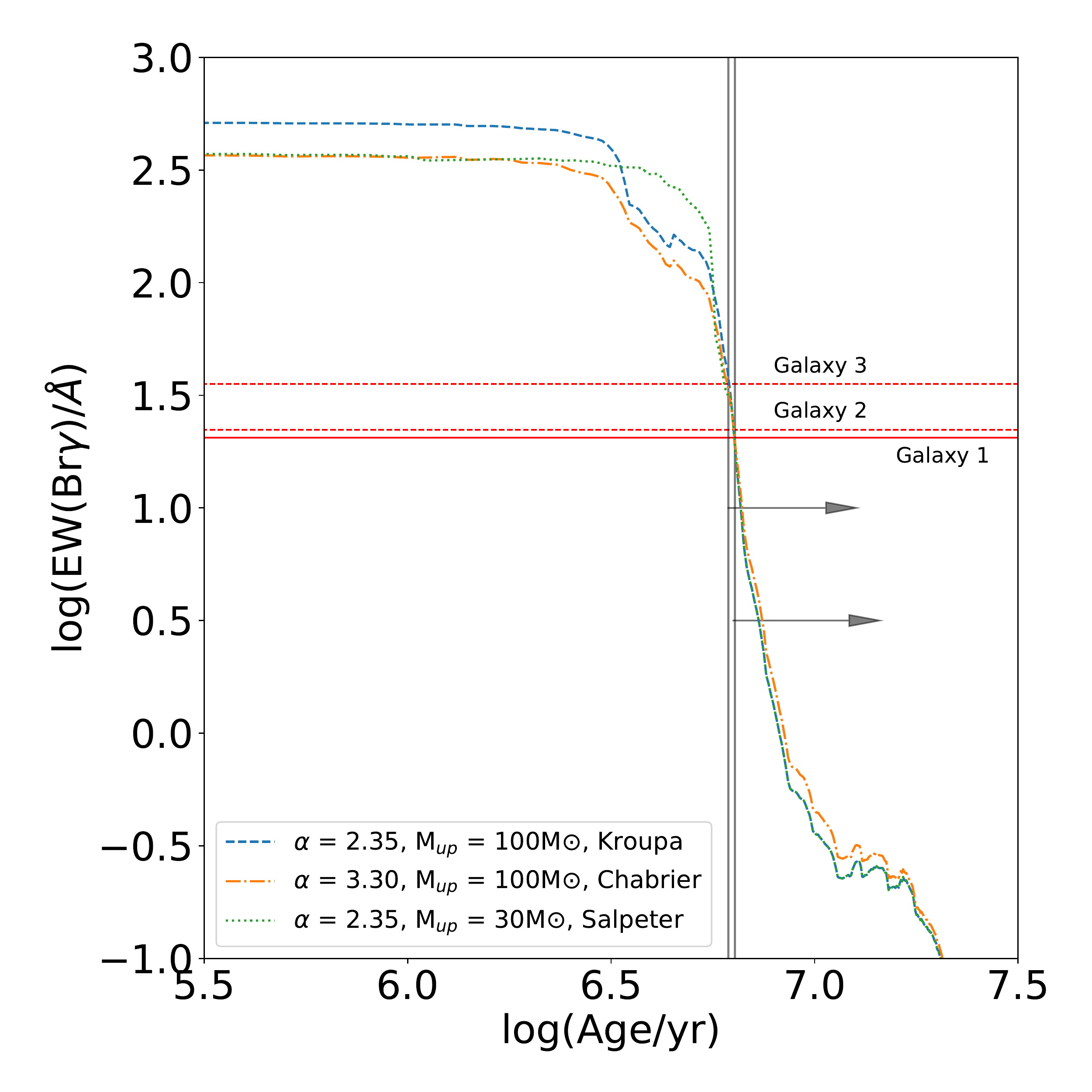}
 \caption{Equivalent width of the Br$\gamma$ emission line from Starburst99 starburst models \citep{Leitherer1999} for three different initial mass functions at solar metallicity. The red horizontal lines represent the observed values measured for each nucleus; we use dotted line styles for Galaxy 2 and 3 to denote that these ages are upper limits. The vertical black lines constrain the ages of the relatively young stellar populations to $\sim$6.1-6.4 Myr, beyond the age at which the HMXB population is expected to peak.
 }
\label{fig:stellage}
\end{figure}

While we have confirmed the AGN in Galaxy 1 via the detection of near-IR coronal lines, Galaxy 2 and 3 require closer examination. If the SFR within Galaxy 2 and 3 were indeed $\sim$30 M$_{\odot}$ yr$^{-1}$ and $\sim$9 M$_{\odot}$ yr$^{-1}$, respectively --- a requirement to explain the observed X-ray emission --- the optical spectra should be dominated by a very young ($<$10 Myr) and blue stellar population, when the population of high mass X-ray binaries (HMXBs) is expected to be high \citep{linden2010}. \textcolor{black}{The most luminous source of X-ray emission in galaxies arises from HMXBs.\footnote{\textcolor{black}{Ultraluminous X-ray Sources (ULXs) can be more luminous than HMXBs, however the vast majority of these objects exhibit unabsorbed $0.2-10$ keV luminosities between $10^{39}-10^{40}$ erg s$^{-1}$ \citep{sutton2012}, which fall below the \textit{absorbed} luminosities of the sources presented here. Additionally, ULXs are not generally associated with mid-infrared AGN colors \citep{secrest2015a}.}}} \textcolor{black}{Our near-infrared spectra allow us to also test this hypothesis by constraining the ages of the nuclear stellar populations. Specifically, the equivalent width of hydrogen recombination lines show a steep decline as the most massive stars evolve off the main sequence, causing a simultaneous decrease in the ionizing photon flux and an increase in the K-band continuum flux \citep{Leitherer1999}.} However, the equivalent width of the Br-$\gamma$ line (or upper limit in the case of Galaxy 2 and 3), a strong indicator of the starburst age \citep{Leitherer1999,Leitherer2014}, suggests that the stellar population in all three galaxies is greater than $\sim$6 million years (Figure~\ref{fig:stellage}), well after the time when the population of HMXBs is expected to drop dramatically \citep{linden2010}.

We also considered the possibility that the optical emission lines are produced by ionizing shocks powered by starburst driven winds or tidal flows \citep{veilleux2002,colina2005}. Shocks are known to arise in mergers and become an increasingly important component of the optical emission lines as a merger progresses \citep{rich2011}, generating [OIII]/H$\beta$ and [NII]/H$\alpha$ ratios consistent with AGN when in fact only star formation is present \citep{allen2008}. However, shock models have shown that when there is a significant contribution from shocks to the ionization, it often results in enhanced [OI]/H$\alpha$ and [SII]/H$\alpha$ emission line ratios compared to AGN photoionization models \citep{dopita1995,allen2008}, which is not seen in any of the three nuclei. Furthermore, detection of three X-ray point sources coincident with the nuclei supports an AGN origin rather than the result of shock-heated gas. 

Finally, we explored the possibility that there are less than three distinct sources of ionization (one located at each galactic nucleus). In particular, given the [OIII] luminosities of each nucleus, the typical narrow line region (NLR) size is expected to be comparable to the nuclear pair separations between Galaxy 1 and Galaxy 2 \citep{bennert2006,hainline2013,hainline2014}, raising the possibility that only one AGN is ionizing the gas at both locations. However the detection of \textcolor{black}{three} separate luminous X-ray sources at the location of each nucleus makes this scenario unlikely. Further, the velocity offset of the [OIII] lines between the two nuclei is consistent with the velocity offset measured for the H$\beta$ lines, suggesting a lack of ionization stratification \citep{zamanov2002} expected for a single AGN, and therefore also favoring a scenario in which there are two independent sources of ionization. Moreover, each galaxy has unique, well-coupled gas and stellar velocity dispersions, suggesting three kinematically distinct sources consistent with galaxy nuclei.

We note that SDSS J0849+1114 was targeted for multiwavelength follow-up as a result of mid-infrared color preselection  \citep{satyapal2014,satyapal2017,pfeifle2019}. While optical preselection strategies \citep{comerford2015,liu2011} are extensively used in campaigns to identify dual AGNs, these strategies have yielded very few confirmed cases. In fact, very few late stage galaxy mergers are known to host dual AGNs (see Table 8 from Satyapal et al. 2017), most of which have been discovered serendipitously or through hard X-ray surveys \citep{koss2012}. While there have been a couple of triplet quasars reported in the literature \citep{djorgovski2007,farina2013}, they are at much larger separations and it is unclear if they are hosted in mergers. A few candidate triple AGNs in mergers have been reported \citep{koss2012,liu2011,kalfountzou2017} but thus far there have been no cases exhibiting triple X-ray point sources {\it and} optical narrow line signatures consistent with AGNs. Emission from the nuclei of advanced mergers at these pair separations are expected to be highly absorbed (e.g.~\citealp{kocevski2015,ricci2017mnras,blecha2018}), making it difficult to identify dual AGNs using traditional techniques. While it is true that SDSS J0849+1114 exhibits optical signatures of AGNs in each nucleus, it is unlikely that all triple AGNs will exhibit such optical spectroscopic signatures. In contrast, mid-infrared preselection has proven an effective strategy in finding dual AGNs and now a triple AGN candidate in follow-up multiwavelength studies \citep{satyapal2017,ellison2017,pfeifle2019}, suggesting that similar cases to SDSS J0849+1114 can be found in future studies. 

\section{Conclusion}
\textcolor{black}{In this investigation, we employed a suite of multiwavelength diagnostics which yield a compelling case for the existence of an AGN triplet in the triple galaxy merger SDSS J0849+1114. This merger, preselected in our dual AGN campaign for follow-up observations based on its \WISE{} mid-infrared colors, was only later realized as a triple AGN candidate \citep{pfeifle2019}. The nature of SDSS J0849+1114 can be summarized as follows:}
\textcolor{black}{
\begin{enumerate}
    \item \chandra{} revealed three nuclear X-ray point sources which exhibit luminosities in excess of any expected contributions from stellar activity.
    \item High excitation [SiVI] coronal line emission was detected in the nucleus of Galaxy 1 via the near-infrared longslit spectra obtained with the LBT. \textcolor{black}{We also find evidence for broad components in the near-IR Pa$\alpha$ emission of Galaxy 1 and 3, with significantly greater widths than those observed in the broad components of the optical [OIII]$\lambda5007$ emission.}
    \item New longslit optical spectroscopic measurements shown for the first time here for Galaxy 1 and Galaxy 2, along with the archival \textit{SDSS} spectrum for Galaxy 3, reveal optical spectroscopic emission line ratios consistent with AGN photoionization in all three nuclei.
    \item Simultaneous spectral fitting of the \chandra{} and new \nustar{} observations confirms at least one of the nuclei (Galaxy 1) hosts a heavily absorbed AGN and we further refine our previously reported spectral parameters: log($\nh{}$/cm$^{-2}$) $=23.9^{+0.2}_{-0.2}$, \textcolor{black}{photon index} $\Gamma=1.4^{+0.3}_{\rm{NC}}$, scattered fraction $f_S=11.7^{+22.1}_{-6.4}$\%, and Fe K$\alpha$ equivalent width upper limit of $\sim0.6$ keV. We find an unabsorbed $2-10$ keV luminosity of $4.6^{+0.6}_{-0.6}\times10^{42}$ erg s$^{-1}$.
    \item Based upon our multiwavelength diagnostics, alternative scenarios --- such as star formation activity, shock-driven emission, or photoionization by fewer than three AGNs --- cannot effectively explain our observations.
\end{enumerate}
}
\textcolor{black}{
Follow-up radio observations of SDSS J0849+1114 obtained with Very Long Baseline Interferometry (VLBI) are presented in Gabanyi et al., (in preparation).}

\textcolor{black}{The presence of triple AGNs in hierarchical galaxy mergers is a natural consequence of $\Lambda$CDM predictions. Since late stage mergers are predicted to be highly obscured \citep{hopkins2008a,blecha2018}, the apparent dearth of dual and triple AGNs  may in fact be due to the pre-selection strategy adopted to find them. With the serendipitous identification of one compelling case for a triple AGN originally selected using our \WISE{} preselection strategy, the natural next step is to extend our study to identify further cases of triple AGNs using mid-infrared color selection. In a forthcoming investigation, we hope to examine a sample of \WISE{} selected ($W1[3.4\mu\rm{m}]-\textit{W}2[4.6\mu\rm{m}]>0.5$) triple mergers in a systematic search for triple AGNs using an arsenal of optical, near-infrared, and X-ray diagnostics.}\\

\acknowledgements
R.\init W.\init P. and S.\init S.~gratefully acknowledge support from the \textit{Chandra} Guest Investigator Program under NASA Grants GO6-17096X and GO7-18099X. R.W.P thanks the \chandra{} Help Desk for support during this project. \textcolor{black}{R.W.P. thanks P. Boorman for the helpful discussions and assistance in troubleshooting the \nustar{} pipeline while R.W.P. became acquainted with the software.} C.\init M.-K., R.\init O.\init S., and G.\init C. acknowledge support from the National Science Foundation, under grant number AST 1817233. C.\init R. acknowledges support from the CONICYT+PAI Convocatoria Nacional subvencion a instalacion en la academia convocatoria a\~{n}o 2017 PAI77170080. \textcolor{black}{L.\init B.\init acknowledges support from the National Science Foundation under grant number AST-1715413.}

\textcolor{black}{This research made use of APLpy, an open-source plotting package for Python \citep{robitaille2012}, as well as the Ned Wright Cosmology Calculator \citep{wright2006}. We also gratefully acknowledge the use of the software TOPCAT \citep{Taylor2005} and Astropy \citep{astropy2013}.} 

The LBT is an international collaboration among institutions in the United States, Italy and Germany. LBT Corporation partners are: The University of Arizona on behalf of the Arizona Board of Regents; Istituto Nazionale di Astrofisica, Italy; LBT Beteiligungsgesellschaft, Germany, representing the Max-Planck Society, The Leibniz Institute for Astrophysics Potsdam, and Heidelberg University; The Ohio State University, and The Research Corporation, on behalf of The University of Notre Dame, University of Minnesota and University of Virginia.

\textcolor{black}{This publication makes use of data products from the Wide-field
Infrared Survey Explorer, which is a joint project of the University
of California, Los Angeles, and the Jet Propulsion
Laboratory/California Institute of Technology, funded by the National
Aeronautics and Space Administration.
Funding for SDSS-III has been provided by the Alfred P. Sloan Foundation, the Participating Institutions, the National Science Foundation, and the U.S. Department of Energy Office of Science. The SDSS-III web site is \url{http://www.sdss3.org/}.}

\textcolor{black}{SDSS-III is managed by the Astrophysical Research Consortium for the Participating Institutions of the SDSS-III Collaboration including the University of Arizona, the Brazilian Participation Group, Brookhaven National Laboratory, Carnegie Mellon University, University of Florida, the French Participation Group, the German Participation Group, Harvard University, the Instituto de Astrofisica de Canarias, the Michigan State/Notre Dame/JINA Participation Group, Johns Hopkins University, Lawrence Berkeley National Laboratory, Max Planck Institute for Astrophysics, Max Planck Institute for Extraterrestrial Physics, New Mexico State University, New York University, Ohio State University, Pennsylvania State University, University of Portsmouth, Princeton University, the Spanish Participation Group, University of Tokyo, University of Utah, Vanderbilt University, University of Virginia, University of Washington, and Yale University.
This research has made use of the NASA/IPAC Extragalactic Database (NED) which is operated by the Jet Propulsion Laboratory, California Institute of Technology, under contract with the National Aeronautics and Space Administration. }

\clearpage 
\begin{appendix}

\renewcommand{\thefigure}{A\arabic{figure}}
\setcounter{figure}{0}

\begin{figure*}[!ht]
    \centering
    \subfloat{\includegraphics[width=1.0\linewidth]{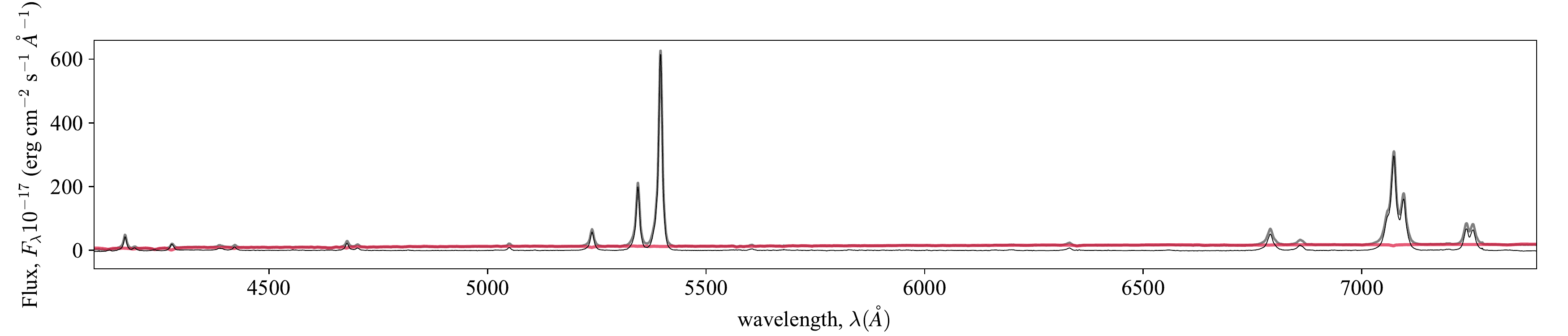}}\\ \vspace{-3mm}
    \subfloat{\includegraphics[width=1.0\linewidth]{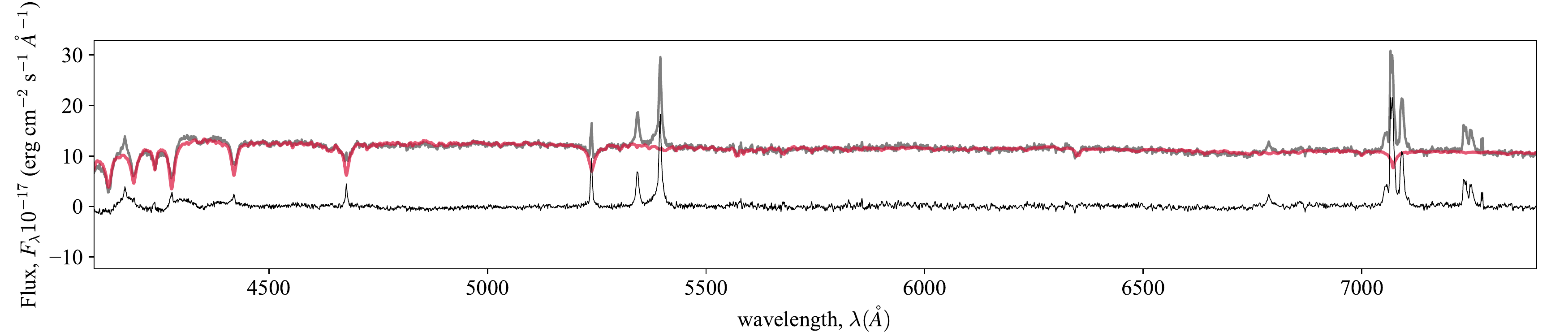}}\\ \vspace{-3mm}
    \subfloat{\includegraphics[width=1.0\linewidth]{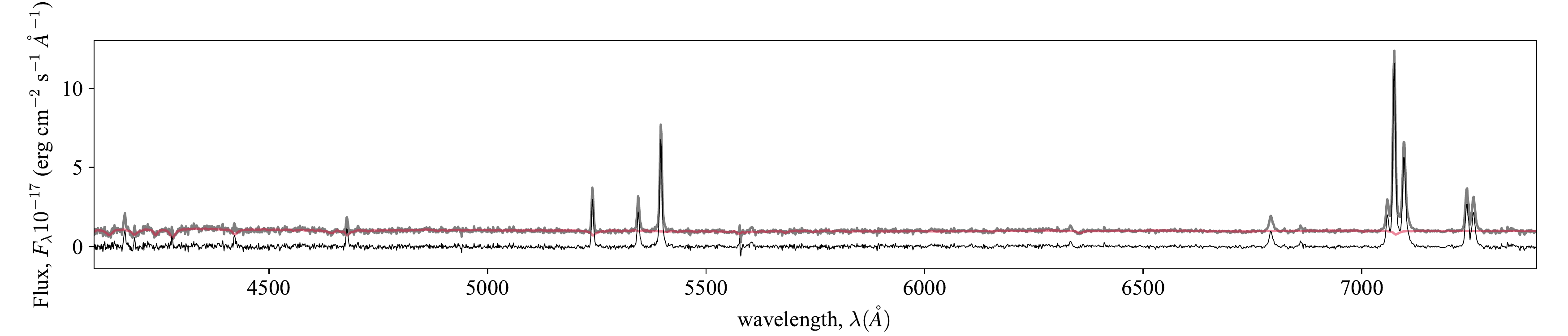}}\\ \vspace{-3mm}
\caption{\textcolor{black}{This figure shows the full dynamic range of the optical spectra shown in Figure~\ref{fig:opticallinefits}. Here, the top, middle, and bottom panels correspond to the spectra for Galaxy 1, 2, and 3, respectively. The spectra for Galaxy 1 and 2 were obtained with the LBT while the spectrum for Galaxy 3 was obtained by the \textit{SDSS}.} The spectra were fit using the procedure outlined in Section 2.6.1; the light gray spectrum in each panel shows the observed spectrum, the red line illustrates the spectrum of the best-fitting stellar population, and the black spectrum of each panel shows the resulting spectrum after subtracting the stellar population model.}
\label{fig:fullopticallinefits}
\end{figure*}

\end{appendix}

\clearpage 

\bibliographystyle{yahapj}

\end{document}